\documentclass[twocolumn,aps]{revtex4}
\newcommand{\beq}{\begin{eqnarray}}
\newcommand{\eeq}{\end{eqnarray}}
\begin{document}
\title{SN1987A Pulsar Velocity From Modified URCA Processes and Landau Levels}
\author{Leonard S Kisslinger, Department of Physics, Carnegie Mellon 
University, Pittsburgh, PA 15213\\ 
Sandip Pakvasa, Department of Physics and Astronomy, University of Hawaii at 
Manoa, Honolulu, HI 96822}

\begin{abstract} Using a recent estimate of the velocity of pulsars arising
from neutrinos emitted with modified URCA processes with electrons in
Landua levels, and the temperture of the protoneutron star created by 
SN1987A, derived from the energy of the observed neutrinos,
we predict the velocity of the resulting pulsar.

\end{abstract}
\maketitle
\noindent
PACS Indices:97.60.Bw,97.60.Gb,97.60.Jd
\vspace{1mm}

   Electrons in very strong magnetic fields, such as those found at the
surface of a protoneutron star created by a supernova, are in Landua levels
\cite{jl,mo}. If the electron is in the lowest Landau level, n=0, it has
only negative helicity with respect to the direction of the magnetic field,
say in the z direction. It was shown in a recent work on pular 
kicks\cite{hjk07} that if electrons created by the modified URCA processes,
which dominate neutrino emission after 10 seconds\cite{bw},
are in the n=0 level, only those moving in the z (B) direction will contribute 
to neutrino emision, and the emitted neutrinos (antineutrinos) are correlated
in the z direction. Therefore, even though only one or two percent of the
neutrino emissivity occurs during the period of approximately 10 to 20 seconds
after the supernova collapse, since almost all emission is correlated
in the z direction this can account for the observed large pulsar velocities:
the pulsar kicks. The resulting pulsar velocity,$ v_{ns}$, is proportional to 
the temperature, T, of the protoneutron star surface to the seventh power,
and one must be able to estimate T in order to predict $ v_{ns}$.

  In the 10 second period in which the modified URCA process dominates   
neutrino emission, the radius of the neutrino sphere, $R_\nu$, is a little
smaller that the radius of the protoneutron star, $R_{ns}$, so all the created
neutrinos correlated with the z direction are emitted.
In ref\cite{hjk07} it was shown that the momentum given to the neutron star
during this period is
\beq
\label{1}
   p_{ns} &\simeq& 0.43 \times 10^{27} P(0)(\frac{T}{10^9 K})^7 
\nonumber \\
        &&(R_{ns}^3-(R_\nu)^3) {\rm \;gm\;cm\;s^{-1}} \; ,
\eeq
where P(0) is the probability of the electron being in the n=0 Landau level.
For the conditions expected in this 10 second period, it was found that
$P(0)\simeq 0.4$, and that $R_\nu \simeq 9.96$ km. Therefore for a neutron
star with the mass of our sun:
\beq
\label{2}
  v_{ns} &=& 1.03 \times 10^{-4} (\frac{T}{10^{10} K})^7 {\rm km\;s^{-1}} \; . 
\eeq
  
   Clearly for a prediction of the velocity of the pulsar, one must know
the temperaure at the surface of the protoneutron star of the protoneutron
quite accurately, since $v_{ns}$ depends on T to the seventh power. If one
knows the energy of the emitted neutrinos at 10 seconds, T is determined by 
the relationship that $kT=E_\nu/3.15$.

   Twenty neutrinos from SN1987A were detected by Kamiokande-II\cite{hir}
and IMB\cite{bio}. The energies of the neutrinos measured by IMB were two
to three times larger than those of Kamiokande-II. This has been discussed 
in many papers. For the present work we need an analysis of the data to
obtain a mean energy of the neutrinos at about 10 seconds. An early model
\cite{bpps87} chose T=$4.1^{+1.0}_{-0.4}$.  Since then there have been
many analyses.  See references \cite{smir,raff}. Although there
are discrepancies, the general agreement is that the neutrino energy at
10 seconds is in the range 9-14 Mev, giving a temperature range:
\beq
\label{3}
             T &\simeq& (3 \leftrightarrow 4.5) {\rm \;\;MeV}=
(3.5 \leftrightarrow 5.2) \times 10^{10} K \; , 
\eeq
which results in our prediction from Eq.(\ref{2}) that
\beq
\label{4}
     v_{ns} &\simeq& (0.6 \rightarrow 10.5) {\rm \;km\;s^{-1}} \;,
\eeq
which is too small in comparison with other sources of pulsar kicks to be 
significant. Note that if the neutrino energy were 30 MeV, $ v_{ns}$ would
be greater than 1000 km/s, observed for high-luminoscity pulsars.

   Another source of pulsar velocities is sterile neutrino oscillation
in the first 10 seconds, with a strong magnetic field resulting in a
significant probability for the electron to be in the n=0 Landau level.
Estimates based on dark matter being sterile neutrinos\cite{fkmp03}, and 
the recent MiniBooNE experiment\cite{mini} analyzed with the LSND 
results in terms of two light sterile neutrinos\cite{ms07} for the range 
of possible pulsar velocities\cite{khj07}, find that sterile neutrinos can 
give pulsar velocities in the observed 1000 km/s range. Since sterile 
neutrino energies cannot be measured by neutrino detectors, however, it is 
not possible to predict the velocity of a pulsar resulting from a supernova 
event.

\end{document}